\documentclass[epj]{svjour}
\usepackage{epsfig}

\begin{document}

\title{A MICROSCOPIC $NN \to NN^*(1440)$ POTENTIAL}
\author{B. Juli\'a-D\'\i az\inst{1,2}, A. Valcarce\inst{2}, 
P. Gonz\'alez\inst{3} \and F. Fern\'andez\inst{2}}

\institute{Department of Physical Sciences, University of Helsinki and
Helsinki Institute of Physics,
P.O. Box 64, 00014 Helsinki, Finland \and
Grupo de F\' \i sica Nuclear,
Universidad de Salamanca, E-37008 Salamanca, Spain \and
Dpto. de F\' \i sica Te\'orica and IFIC,
Universidad de Valencia - CSIC, E-46100 Burjassot, Valencia, Spain}

\date{Received: date / Revised version: date}
%
\abstract{By means of a $NN \to NN^*(1440)$ transition
potential derived in a parameter-free way from a quark-model
based $NN$ potential,
we determine simultaneously the $\pi NN^*(1440)$ and $\sigma NN^*(1440)$
coupling constants. We also present a study of the target Roper excitation
diagram contributing to the $p(d,d')$ reaction.
 } 
\maketitle
\section{Introduction}

The $N^{\ast }(1440)$ (Roper) is a broad resonance which couples strongly (60%
$-$70$\%$) to the $\pi N$ channel and significantly (5$-$10$\%$) to the $%
\sigma N$ channel \cite{14}. These features suggest that the Roper resonance
should play an important role in nuclear dynamics as an intermediate state.
Graphs involving the excitation of $N^{\ast }(1440)$ appear in different
systems, as for example the three-nucleon interaction mediated by $\pi $ and 
$\sigma $ exchange contributing to the triton binding energy \cite{2}. The
excitation of the Roper resonance has also been used to explain the missing
energy spectra in the $p(\alpha,\alpha^{\prime})$ reaction \cite{10} or the $%
np\to d(\pi\pi)^0$ reaction \cite{3}. The coupling of the $N^{\ast }(1440)$
to $\pi N$ and $\sigma N$ channels could also be important in heavy ion
collisions at relativistic energies \cite{4,12}. Finally, pion electro- and
photoproduction may take place through the $N^*(1440)$ excitation \cite{1}.
However the use of a $NN\rightarrow NN^{\ast }(1440)$ transition potential
as a straightforward generalization of some pieces of the $NN\rightarrow NN$
potential plus the incorporation of resonance width effects may have serious
shortcomings specially concerning the short-range part of the interaction 
\cite{7}.

In this talk we present some applications of a recently derived $%
NN\rightarrow NN^{\ast }(1440)$ interaction \cite{5}, obtained 
by means of the same quark-model approach 
previously used to study the $NN$ system and transition
potentials involving the $\Delta$. A main feature of the quark treatment is
its universality in the sense that all the baryon-baryon interactions are
treated on an equal footing. Moreover, once the model parameters are fixed
from $NN$ data there are no free parameters for any other case. This allows
a microscopic understanding and connection of the different baryon-baryon
interactions that is beyond the scope of any analysis based only on
effective hadronic degrees of freedom. These studies are instructive
inasmuch as they are expected to lead to a deeper understanding of the
nuclear potential and entail a rethinking of basic nuclear concepts from the
point of view of the fundamental quark substructure.
We center our attention in the
derivation of the $\pi NN^*(1440)$ and $\sigma NN^*(1440)$ coupling
constants and in the study of a reaction mediated by the excitation of the
Roper resonance, the $p(d,d^{\prime})$ reaction.

\section{$\protect\pi NN^*(1440)$ and $\protect\sigma NN^*(1440)$ coupling
constants.}

The usual way to determine meson$-NN$ coupling constants is trough the
fitting of $NN$ scattering data with phenomenological meson exchange models.
Therefore, a consistent way to obtain meson$-NN^*$ coupling constants
is from a transition $NN \to NN^*$ potential, in particular when
ratios over meson$-NN$ coupling constants are to be considered. In order to
derive the transition potential we shall follow the same quark model
approach previously used for $NN$ scattering \cite{6}. Explicitly, the $%
NN\rightarrow NN^*(1440)$ potential at interbaryon distance $R$ is obtained
by sandwiching the $qq$ potential, $V_{qq}$, between $NN$ and $NN^*(1440)$
states, written in terms of quarks, for all the pairs formed by two quarks
belonging to different baryons. The $qq$ potential contains a confining term
taken to be linear ($r_{ij}$), the usual 
perturbative one-gluon-exchange (OGE)
interaction containing Coulomb ($1/r_{ij}$), spin-spin (${\vec{\sigma}}%
_{i}\cdot {\vec{\sigma}}_{j})$ and tensor ($S_{ij}$) terms, and pion and
sigma exchanges as a consequence of the breaking of chiral symmetry. The 
wave function of the Roper,
$N^*(1440)$, and nucleon, $N$, states can be written as 
$|N^*(1440)\rangle =\left\{ \sqrt{%
\frac{2}{3}}|[3](0s)^{2}(1s)\rangle - \sqrt{\frac{1}{3}}|[3](0s)(0p)^{2}%
\rangle \right\} \otimes \lbrack 1^{3}]_{c}$ and $|N\rangle
=|[3](0s)^{3}\rangle \otimes \lbrack 1^{3}]_{c}$ where $[1^{3}]_{c}$ is the
completely antisymmetric color state, $[3]$ is the completely symmetric
spin-isospin state and $0s$, $1s$, and $0p$, stand for harmonic oscillator
orbitals.

The transition potential obtained can be written at all distances in terms
of baryonic degrees of freedom \cite{8}. One should realize that a $qq$ spin
and isospin independent potential as for instance the scalar 
one-sigma exchange (OSE), gives rise
at the baryon level, apart from a spin-isospin independent potential, to a
spin-spin, an isospin-isospin and a spin-isospin dependent interactions \cite
{5}. Nonetheless for distances $R\geq 4$ fm, where quark antisymmetrization
interbaryon effects vanish, we are only left with the direct part, i.e. with
a scalar OSE at the baryon level. The same kind of arguments can be applied
to the one-pion exchange (OPE) potential. 
Thus asymptotically ($R\geq 4$ fm) OSE and OPE have at
the baryon level the same spin-isospin structure than at the quark level.
Hence we can parametrize the asymptotic central interactions as 
\begin{eqnarray}
V_{NN\rightarrow NN^*(1440)}^{OPE}(R) & = & \frac{1}{3}\,\frac{g_{\pi NN}} {%
\sqrt{4\pi }}\,\frac{g_{\pi NN^{\ast }(1440)}}{\sqrt{4\pi }}\,\frac{m_{\pi} 
}{2M_{N}}\, \nonumber \\ \frac{m_{\pi }}{2(2M_{r})} 
\frac{\Lambda ^{2}}{\Lambda ^{2}-m_{\pi }^{2}} &[&(\vec{\sigma}_{N}.\vec{%
\sigma}_{N})(\vec{\tau}_{N}.\vec {\tau}_{N})]\,\frac{e^{-m_{\pi }R}}{R}\,,
\label{lrg}
\end{eqnarray}
and 
\begin{eqnarray}
V_{NN\rightarrow NN^*(1440)}^{OSE} (R) & = & - \, \frac{g_{\sigma NN}}{\sqrt
{4\pi }} \, \frac{g_{\sigma NN^{\ast}(1440)}}{\sqrt{4\pi }} \,  \nonumber \\
& & \frac{\Lambda^{2}}{\Lambda ^{2}-m_{\sigma }^{2}} \, \frac{e^{-m_{\sigma
}R}}{R} \, ,  \label{slrg}
\end{eqnarray}
where $g_{i}$ stands for the coupling constants at the baryon level and $%
M_{r}$ is the $NN^*(1440)$ reduced mass.

By comparing the baryonic potentials with the asymptotic behavior of the
ones previously obtained from the quark-model calculation we can extract
the $\pi NN^{\ast }(1440)$ and $\sigma NN^{\ast }(1440)$ coupling constants.
As the parameters at the quark level are fixed once for all from the $NN$
interaction our results allow a prediction of these constants in terms of
the elementary $\pi qq$ coupling constant and the one-baryon model dependent
structure. The sign obtained for the meson-$NN^{\ast }(1440)$ coupling
constants and for their ratios to the meson-$NN$ coupling constants is
ambiguous since it comes determined by the arbitrarily chosen relative sign
between the $N$ and $N^{\ast }(1440)$ wave functions. Only the ratios
between the $\pi NN^{\ast }(1440)$ and $\sigma NN^{\ast }(1440)$ would be
free of this uncertainty. This is why we will quote absolute values except
for these cases where the sign is a clear prediction of the model. To get
such a prediction we can use any partial wave. We shall use for simplicity
the $^{1}S_{0}$ wave, this is why we only wrote the central interaction in
Eq.~(\ref{lrg}).

The $[\Lambda ^{2}/({\Lambda ^{2}-m_i^{2}})]$ factor comes from the
vertex form factor chosen at momentum space as a square root of monopole $[
\Lambda ^{2} / ({\Lambda ^{2}+ \vec{q}^{\,\,2}}) ]^{1/2}$, the same choice
taken at the quark level, where chiral symmetry requires the same form for
pion and sigma. A different choice for the form factor at the baryon level,
regarding its functional form as well as the value of $\Lambda $, would give
rise to a different vertex factor and eventually to a different functional
form for the asymptotic behavior. For instance, for a modified monopole
form, $[ (\Lambda ^{2}-m^{2}) / ({\Lambda ^{2}- \vec{q}^{\,\,2}}) ]^{1/2}$,
where $m$ is the meson mass ($m_{\pi }$ or $m_{\sigma }$), the vertex factor
would be $1$, i.e. $[(\Lambda ^{2}-m^{2})/({\Lambda ^{2}-m^{2}})]$, keeping
the potential the same exponentially decreasing asymptotic form. Then it is
clear that the extraction from any model of the meson-baryon-baryon coupling
constants depends on this choice. We shall say they depend on the coupling
scheme.

\begin{figure}[t]
\begin{center}
\mbox{\epsfxsize=80mm\epsffile{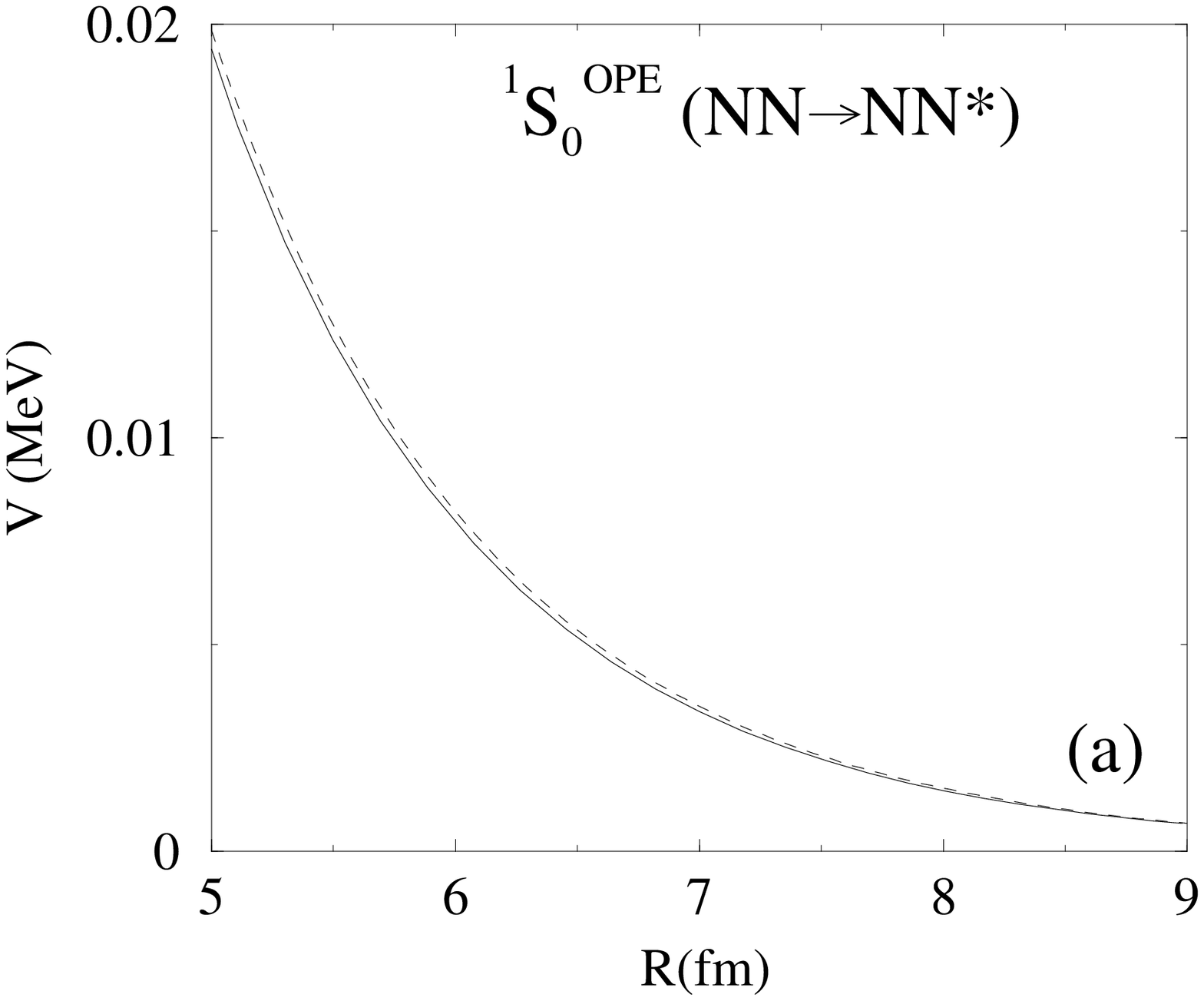}} \mbox{\epsfxsize=80mm%
\epsffile{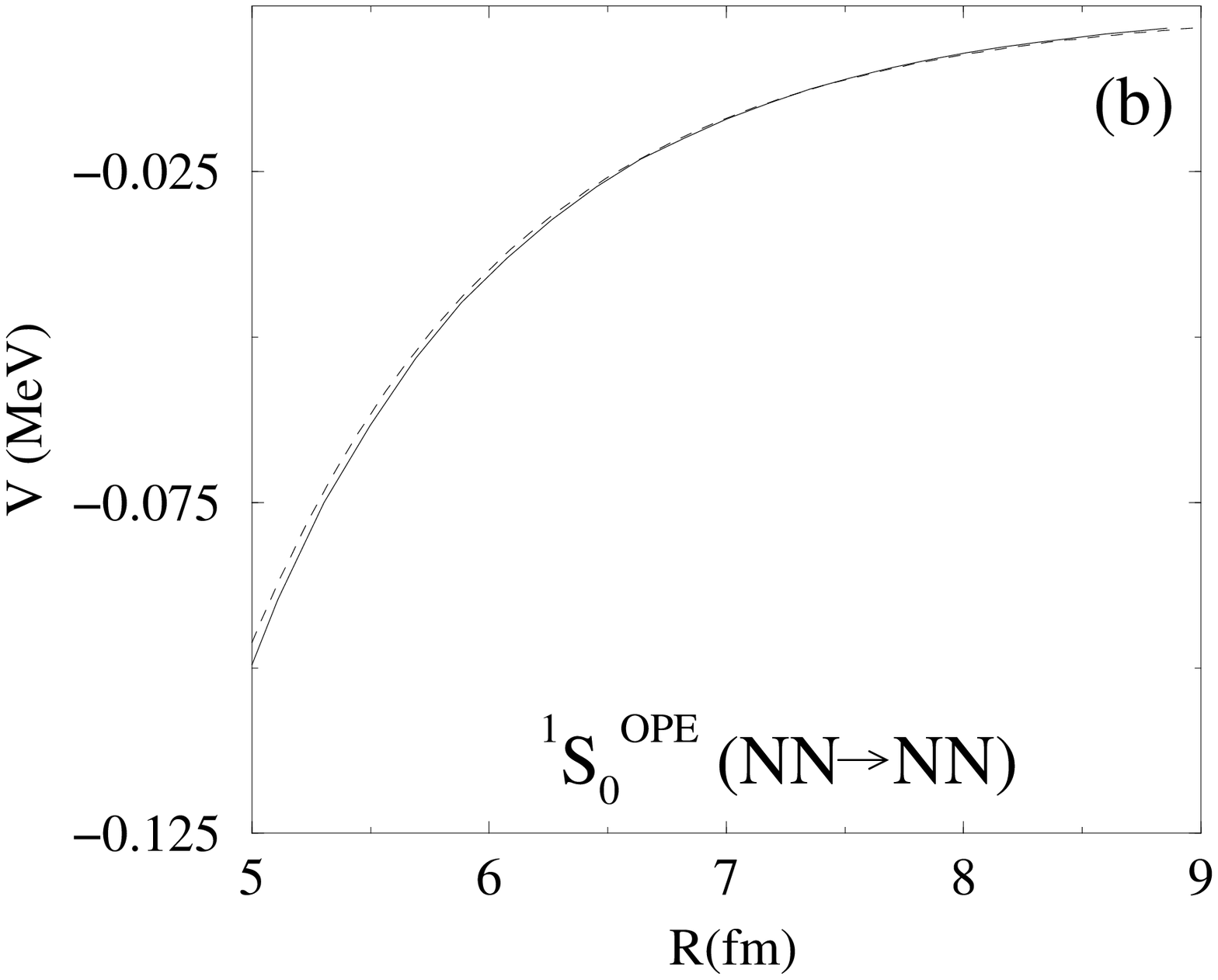}}
\end{center}
\caption{(a) Asymptotic behavior of the one-pion exchange $^1S_0$ $NN \to
NN^*(1440)$ potential (solid line). The dashed line denotes the fitted curve
according to Eq. (\ref{lrg}). (b) Same as (a) but for the one-pion exchange $%
^1S_0$ $NN \to NN$ potential.}
\label{fignr1s0lrgtran}
\end{figure}

\begin{figure}[ht]
\begin{center}
\mbox{\epsfxsize=80mm\epsffile{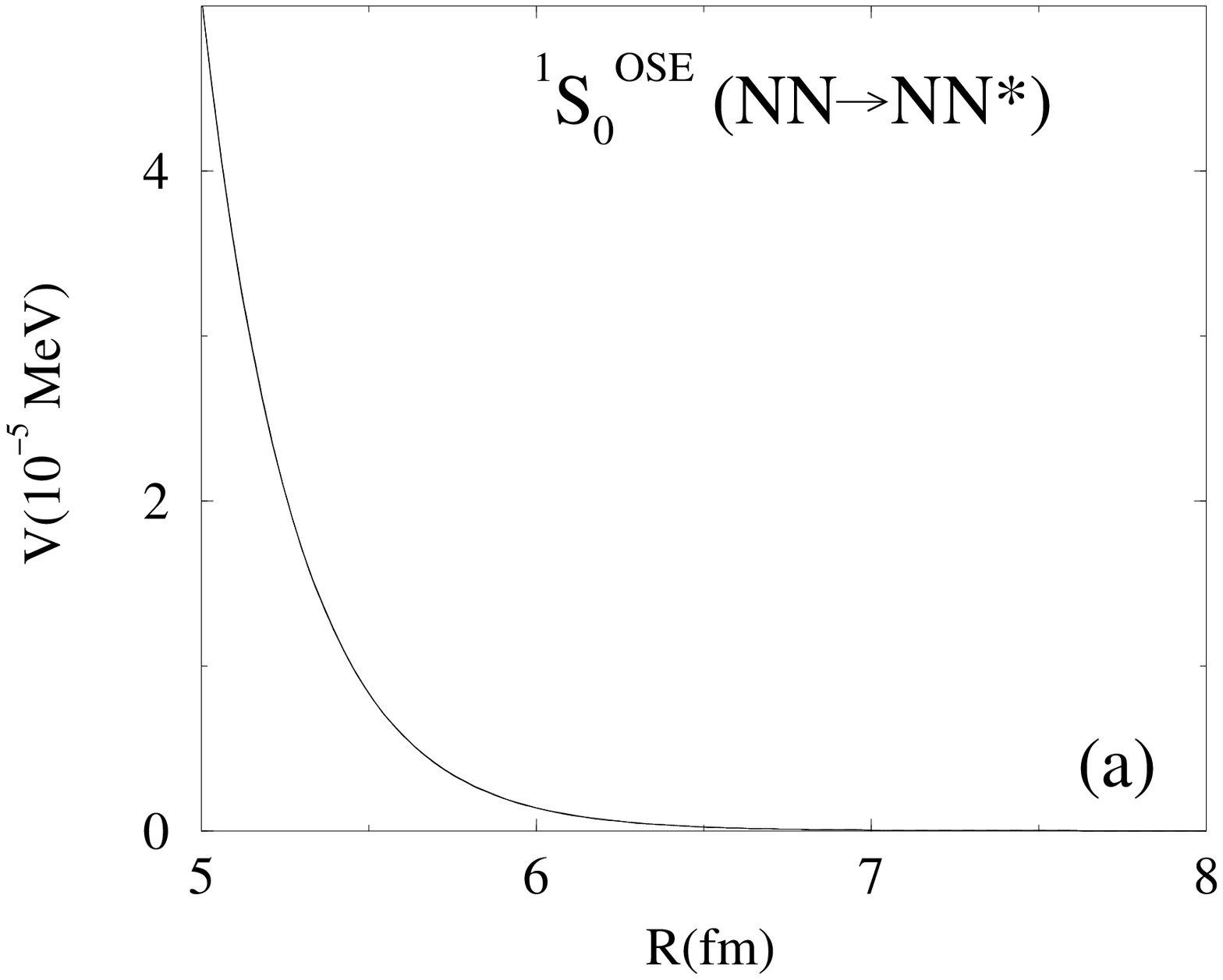}} \mbox{\epsfxsize=80mm%
\epsffile{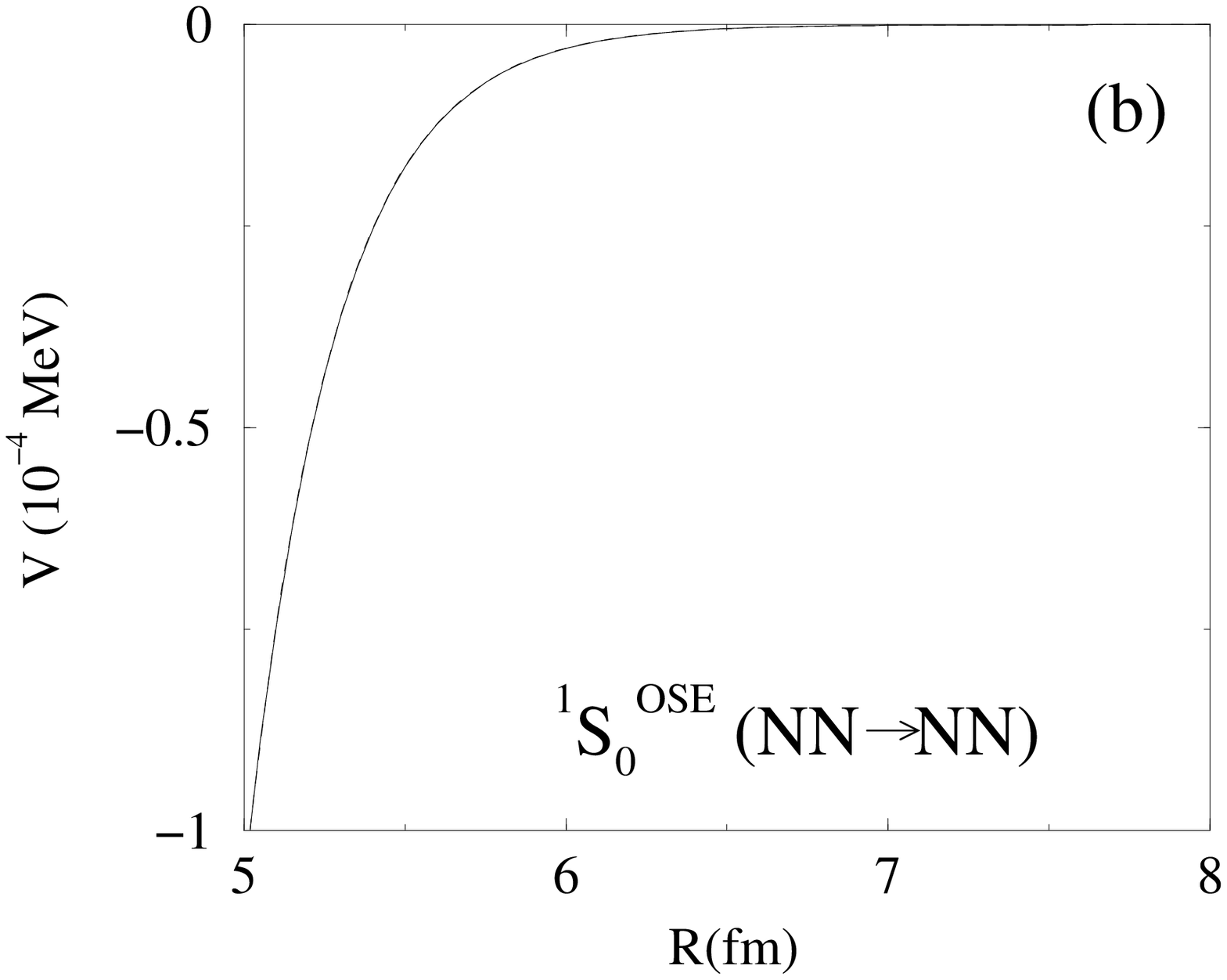}}
\end{center}
\caption{(a) Asymptotic behavior of the one-sigma exchange $^1S_0$ $NN \to
NN^*(1440)$ potential (solid line). The dashed line denotes the fitted curve
according to Eq.~(\ref{slrg}). (b) Same as (a) but for the one-sigma
exchange $^1S_0$ $NN \to NN$ potential.}
\label{fignr1s0silrgtran}
\end{figure}

For the one-pion exchange and for our value of $\Lambda =4.2$ fm$^{-1}$, $%
[\Lambda ^{2}/ (\Lambda ^{2}-m_{\pi }^{2})]=1.03$, pretty close to $1$. As a
consequence, in this case the use of our form factor or the modified
monopole form at baryonic level makes little difference in the determination
of the coupling constant. This fact is used when fixing $g_{\pi qq}^{2}/{%
4\pi }$ from the experimental value of $g_{\pi NN}^{2}/{4\pi }$ extracted
from $NN$ data.

To get $g_{\pi NN^*(1440)}/\sqrt{4\pi }$ we turn to our results for the $%
^{1}S_{0}$ OPE potential, Fig. \ref{fignr1s0lrgtran}, and fit its asymptotic
behavior (in the range $R:5\rightarrow 9$ fm) to Eq. (\ref{lrg}). We obtain 
\begin{equation}
\frac{g_{\pi NN}}{\sqrt{4\pi }} \frac{g_{\pi NN^{\ast}(1440)}}{\sqrt{4\pi }} 
\frac{\Lambda ^{2}}{\Lambda ^{2}-m_{\pi }^{2}}= \, - \, 3.73 \, ,
\end{equation}

\noindent i.e. $g_{\pi NN^{\ast }(1440)}/\sqrt{4\pi }= - 0.94$. As explained
above only the absolute value of this coupling constant is well defined. Let
us note that in Ref. \cite{9} a different sign with respect to our coupling
constant is obtained what is a direct consequence of the different global
sign chosen for the $N^{\ast }(1440)$ wave function. The coupling scheme
dependence can be explicitly eliminated if we compare $g_{\pi NN^{\ast
}(1440)}$ with $g_{\pi NN}$ extracted from the $NN\rightarrow NN$ potential
within the same quark model approximation, Fig. \ref{fignr1s0lrgtran}. Thus
we get 
\begin{equation}
\left | \frac{g_{\pi NN^{\ast}(1440)}}{g_{\pi NN}} \right |=0.25 \,.
\label{eq17}
\end{equation}

By proceeding in the same way for the OSE potential, i.e. by fitting the
potential given in Fig. \ref{fignr1s0silrgtran}(a) to Eq. (\ref{slrg}), and
following an analogous procedure for the $NN$ case, Fig. \ref
{fignr1s0silrgtran}(b), we can write 
\begin{equation}
\left |\frac{g_{\sigma NN^{\ast}(1440)}}{g_{\sigma NN}} \right |=0.47 \, .
\label{eq18}
\end{equation}

The ratio given in Eq. (\ref{eq17}) is similar to that obtained in Ref. \cite
{9} and a factor 1.5 smaller than the one obtained from the analysis of the
partial decay width. Nonetheless one can find in the literature values for $%
f_{\pi NN^{\ast }(1440)}$ ranging between 0.27$-$0.47 coming from different
experimental analyses with uncertainties associated to the fitting of
parameters \cite{3,12,1}.

Regarding the ratio obtained in Eq. (\ref{eq18}), our result agrees quite
well with the only experimental available result, obtained in Ref. \cite{11}
from the fit of the cross section of the isoscalar Roper excitation in $%
p(\alpha,\alpha^{\prime})$ in the 10$-$15 GeV region, where a value of 0.48
is given. Furthermore, we can give a very definitive prediction of the
magnitude and sign of the ratio of the two ratios,

\begin{equation}
\frac{g_{\pi NN^{\ast}(1440)}}{g_{\pi NN}}=0.53 \; \frac{g_{\sigma
NN^{\ast}(1440)}}{g_{\sigma NN}} \, ,
\end{equation}
which is an exportable prediction of our model.

\section{Roper excitation in \lowercase{$pd$} scattering}

\begin{figure}[b]
\begin{center}
\mbox{\epsfig{file=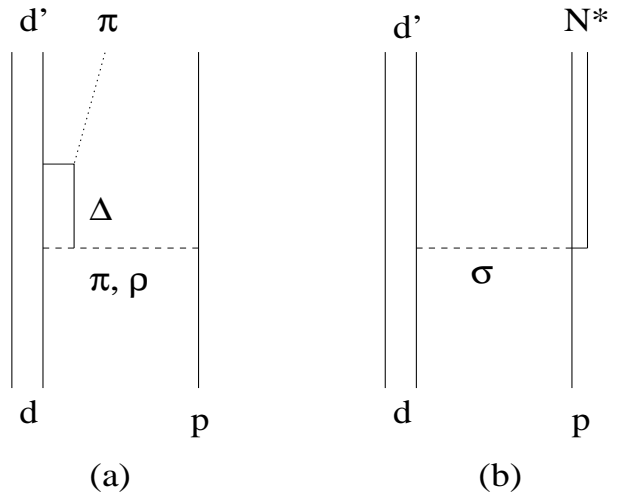, width=80mm, height=65mm}}
\end{center}
\par
\centering
\caption{Dominant mechanisms contributing to 
the $p(d,d^{\prime})$ reaction \protect\cite
{17}.}
\label{fig:hiremec}
\end{figure}

There are two experiments where the contribution from the $N^*(1440)$
resonance has been isolated by 
means of model-dependent theoretical methods. The
first one is the $p (\alpha,\alpha^{\prime})$ reaction carried out in Saclay 
\cite{15} already ten years ago. The data showed two peaks in the cross
section that were not understood for some years. The most prominent one was
attributed to a $\Delta$ excitation in the projectile (DEP) \cite{16}. The
second peak was explained when a Roper excitation in the target (RET) was
considered \cite{10} giving a plausible explanation to the measured
differential cross section.

The second experiment is the $p (d,d^{\prime})$ reaction. It was considered
and studied making use of the same mechanisms \cite{17}. In Fig. \ref
{fig:hiremec} we show the two diagrams which give the bulk contribution to
the cross section of the processes. 

These two reactions are particularly interesting because in both cases the
projectile ($d$ or $\alpha$) has $T=0$. This ensures that the $N^*(1440)$
reaction mechanism, Fig.~\ref{fig:hiremec}(b), can only be driven by a
scalar interaction. Therefore these reactions have provided a method to
determine the baryonic coupling constant between the $N$, $N^*(1440)$ and
the $\sigma$ meson once the 
$\Delta$ contribution has been fixed. The results for the coupling constants
obtained in this way from $p(\alpha ,\alpha ^{\prime })$ have been quoted
and compared to ours in the previous section. Our purpose in this section is
the study of the $p(d,d^{\prime })$ process making use of the quark model
$NN\rightarrow NN^{\ast }(1440)$ transition potential,
to explore the mechanism proposed in Ref. \cite{17}. The differential
cross section for the process is given by: 
\begin{eqnarray}
&&{\frac{d^{2}\sigma }{dE_{d^{\prime }}d\Omega _{d^{\prime }}^{L}}}={\frac{%
p_{d^{\prime }}}{(2\pi )^{5}}}{\frac{M_{d}^{2}M^{2}}{\lambda
^{1/2}(s,M^{2},M_{d}^{2})}}\times   \nonumber \\
&&\int {\frac{d^{3}p_{\pi }}{E_{N^{\prime }}\omega _{\pi }}}\bar{\Sigma}%
\Sigma |T|^{2}\delta (E_{d}+E_{N}-E_{d^{\prime }}-E_{N^{\prime }}-\omega
_{\pi })\,,  \label{eq:cs2}
\end{eqnarray}
where $M (M_{d}$) is the nucleon (deuteron) mass, $s$ is the invariant mass
of the $p-d$ system, 
$\lambda (x,y,z)=x^{2}+y^{2}+z^{2}-2xy-2yz-2xz$ and 
$\bar{\Sigma }\Sigma |T|^{2}$ is the amplitude for the 
elementary process of $N^{\ast}(1440)$ production. 
This amplitude can be written in terms of the scalar
transition potential $(V_0)_{NN\rightarrow NN^{\ast }}$ \cite{17}: 
\begin{equation}
\bar{\Sigma}\Sigma |T|^{2}=12F_{d}^{2}\left( {\frac{f^{\prime }}{m_{\pi }}}%
\right) ^{2}|G^{\ast }|^{2}\left| (V_0)_{NN\rightarrow NN^{\ast
}}(q_{cm})\right| ^{2}q_{cm}^{2}\,.
\end{equation}
The function $F_{d}(\vec{k})$ is the deuteron form factor defined as 
\begin{equation}
F_{d}(\vec{k})=\int d\vec{r}\;\phi ^{\ast }(\vec{r})\;e^{i{\frac{\vec{k}%
\cdot \vec{r}}{2}}}\;\phi (\vec{r})
\end{equation}
where $\phi (\vec{r})$ is the deuteron S-wave function, and the momentum $%
\vec{k}=\vec{p}_{d}-\vec{p}_{d^{\prime }}$ is taken in the initial deuteron
rest frame. $q_{cm}$ is the momentum transfer between the nucleons
in the center of mass system and
$f^{\prime }\equiv f_{\pi NN^{\ast }}$. $G^{\ast }$ 
is the $N^{\ast}(1440)$ propagator as given in Ref. \cite{17}.

We evaluate the cross section in the center of mass system and then relate
the result to the one which is shown by the experimentalists making use of: 
\begin{equation}
{\frac{d^{2}\sigma }{dE_{d^{\prime }}d\Omega _{d^{\prime }}^{L}}}={\frac{%
d^{2}\sigma }{dE_{d^{\prime }}d\Omega _{d^{\prime }}^{cm}}}\;{\frac{d\Omega
_{d^{\prime }}^{cm}}{d\Omega _{d^{\prime }}^{L}}}\,.
\end{equation}
For the kinematics considered it can be shown that 
\begin{equation}
{\frac{d\Omega _{d^{\prime }}^{cm}}{d\Omega _{d^{\prime }}^{L}}}={\frac{%
p_{d}^{L}p_{d^{\prime }}^{L}}{p_{d}^{cm}p_{d^{\prime }}^{cm}}}\left( 1-{%
\frac{E_{d}^{cm}}{\sqrt{s}}}\right) +{\frac{\cos (\theta ^{cm})}{%
p_{d^{\prime }}^{cm\,2}}}{\frac{E_{d^{\prime }}^{cm}}{\sqrt{s}}}%
p_{d}^{L}p_{d^{\prime }}^{L}\;.
\end{equation}

\begin{figure}[t]
\begin{center}
\vspace{18pt} \mbox{\epsfxsize=80mm \epsfysize=70mm\epsffile{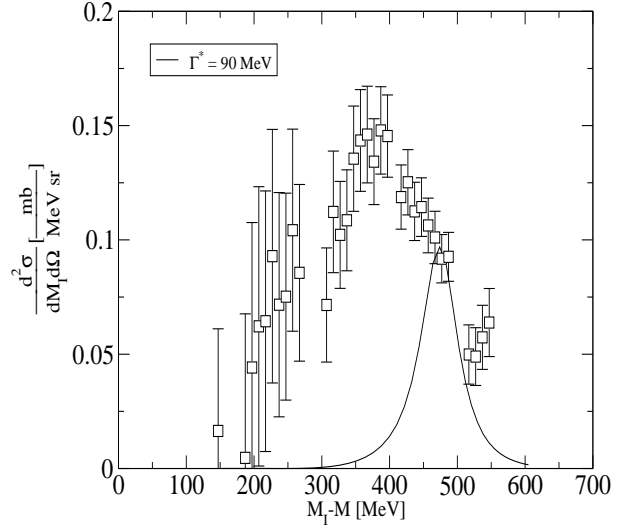}}
\end{center}
\par
\centering
\caption{Quark model result for the RET process contributing to the
$p(d,d')$ reaction. 
$M_I$ is the invariant mass of the target system.
Experimental data correspond to $T_d=$ 2.3 GeV
and $\theta^L=$ 1.1 deg. They were obtained in Ref. \protect\cite{17}
by means of a theoretical subtraction of the $\Delta$ contribution.}
\label{fighire}
\end{figure}

In order to perform the calculation using our quark model, we need to
extract the genuine scalar potential at all distances from our
$NN\rightarrow NN^{\ast }(1440)$ transition potential. At short distances, 
$R<2$ fm, the quark model based potential has a non-trivial structure. Due to
the presence of the antisymmetrizer we have that, for instance, a scalar
coupling at quark level gives rise to a scalar, spin-spin, pseudoscalar and
pseudovector couplings \cite{18}. The extraction of the scalar part can be
done once the unprojected potential for different $(ST)$ channels has been
evaluated in the form:
\begin{eqnarray}
V^{(S,T)}_{NN\rightarrow NN^*} (\vec{q}) &=&
{\frac{4\pi }{\mathcal{N}^{ST}}}\sum_{JM_{J}}^{J_{MAX}}%
\sum_{LM_{L}}\sum_{M_{S}}C_{JM_{J}}^{LSM_{L}M_{S}}Y_{LM_{L}}(\hat{q}) 
\nonumber \\
&&\int_{0}^{\infty }\;r^{2}\;dr\;j_{L}(qr)\;
\hat{V}_{NN\rightarrow NN^*}^{L,S;JT}(r)\,,
\end{eqnarray}
where we are adding up the partial waves up to a certain $J_{MAX}$. $%
\mathcal{N}^{ST}$ is the unprojected norm of the $NN^{\ast }(1440)$ system
and $\hat{V}^{L,S;JT}$ is the projected potential multiplied by the norm of
the corresponding partial wave.

\begin{figure}[b]
\begin{center}
\vspace{18pt} \mbox{\epsfxsize=80mm \epsfysize=70mm\epsffile{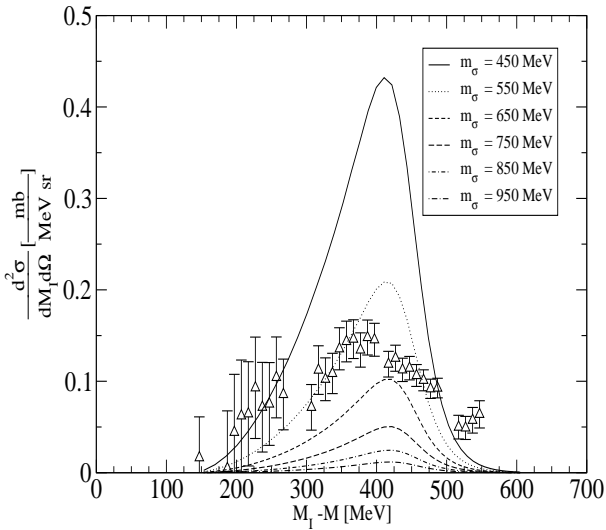}}
\end{center}
\par
\centering
\caption{ Dependence of the $p(d,d')$ cross section 
calculated in Ref. \protect\cite{17} on the
mass of the $\protect\sigma $ meson.
$M_I$ is the invariant mass of the target system.
Experimental data correspond to $T_d=$ 2.3 GeV
and $\theta^L=$ 1.1 deg. They were obtained in Ref. \protect\cite{17}
by means of a theoretical subtraction of the $\Delta$ contribution.}
\label{hireWID1}
\end{figure}

Then we write down the most general form for the interaction: 
\begin{eqnarray}
V_{NN\rightarrow NN^{\ast }}^{(S,T)} &=&V_{0}+
V_1\;\vec{\sigma _{1}}\cdot \vec{%
\sigma _{2}}+V_{2}\;\vec{\tau _{1}}\cdot \vec{\tau _{2}}+  \nonumber \\
&&V_{3}\;\vec{\sigma _{1}}\cdot \vec{\sigma _{2}}\;\vec{\tau _{1}}\cdot \vec{%
\tau _{2}}\,.
\end{eqnarray}
where $V_{i}$ are functions of the interbaryon momentum, ${\vec{\sigma}_{i}}$
and $\vec{\tau}_{i}$ are spin and isospin matrices of the baryons. $V_{0}$
is the scalar part of the total potential which is the only part that can be
included in our process of $N^{\ast }(1440)$ excitation in $p(d,d^{\prime })$
reactions.

Finally, if we consider different $(ST)$ channels, we obtain the 
following system of equations,
\begin{eqnarray}
V_{NN\rightarrow NN^{\ast }}^{(0,0)} &=&V_{0}-3V_{1}-3V_{2}+9V_{3}  \nonumber
\\
V_{NN\rightarrow NN^{\ast }}^{(1,0)} &=&V_{0}+V_{1}-3V_{2}-3V_{3}  \nonumber
\\
V_{NN\rightarrow NN^{\ast }}^{(0,1)} &=&V_{0}-3V_{1}+V_{2}-3V_{3}  \nonumber
\\
V_{NN\rightarrow NN^{\ast }}^{(1,1)} &=&V_{0}+V_{1}+V_{2}+V_{3}\,,
\end{eqnarray}
and solving for the scalar part, 
\begin{eqnarray}
V_{0} &=&{\frac{1}{16}}\left[ V_{NN\rightarrow NN^{\ast
}}^{(0,0)}+3V_{NN\rightarrow NN^{\ast }}^{(0,1)}+\right.   \nonumber \\
&&\left. 3V_{NN\rightarrow NN^{\ast }}^{(1,0)}+9V_{NN\rightarrow NN^{\ast
}}^{(1,1)}\right] \,.
\end{eqnarray}

\begin{figure}[b]
\begin{center}
\vspace{18pt} \mbox{\epsfxsize=80mm \epsfysize=70mm\epsffile{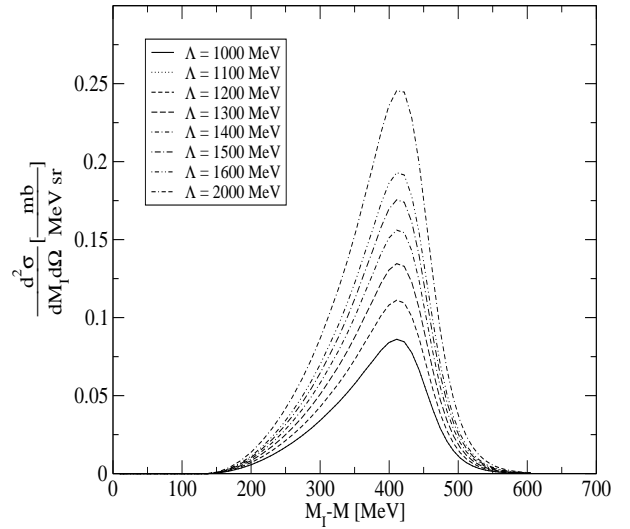}}
\end{center}
\par
\centering
\caption{ Dependence of the $p(d,d`)$ cross section calculated in 
Ref. \protect\cite{17} on the baryonic cut-off mass.
$M_I$ is the invariant mass of the target system.}
\label{hireWID2}
\end{figure}

We focus our attention on the target Roper excitation process. To compare to
data it is necessary to subtract the $\Delta$ contribution and the 
interference term from the
experimental points. The parameters of the $\Delta $ excitation on the
projectile used in the phenomenological model were settled in the ($\alpha
,\alpha ^{\prime }$) reaction. We assume this process to be correctly
described. Therefore we consider the data where the $\Delta $ contribution
has already been subtracted \cite{17} as our experimental data. 

In Fig. \ref{fighire} we show the result obtained using the quark-model
derived $NN\to NN^*(1440)$ potential and its comparison
to data. As can be seen,
the predicted cross section is smaller that the model-dependent experimental
data. If we choose a small value for the width of the $N^*(1440)$, the
results come closer to the experimental data. Let us notice that the bigger
disagreement with the extracted data corresponds to the region where the
error bars are larger, in other words, to the region where the uncertainties
related to the theoretical method used to subtract the $\Delta$ contribution
and interference term are important. For the sake of clarity, let us note
that the subtraction of the $\Delta$ contribution is proportional to the
square of the $\pi N\Delta$ coupling constant. This coupling constant is
different as used in baryonic processes, $f^2_{\pi N\Delta}/4 \pi=0.35$, as
the one used in our quark model, $f^2_{\pi N\Delta}/4 \pi=0.22$ \cite{19}.
This value is crucial when trying to reproduce the $^1S_0$ $NN$ phase shift
through the tensor coupling to the $^5D_0$ $N\Delta$. Using the baryonic
coupling one would obtain much bigger attraction than observed
experimentally. As a consequence, the baryonic calculation of the $\Delta$
contribution could be underestimating the region above the peak
overestimating in this way the $N^*(1440)$ contribution. The way to wipe out
those uncertainties would be to calculate the $\Delta$ contribution together
with the interference term making use of quark-model baryonic potentials.

\begin{figure}[t]
\begin{center}
\vspace{15pt} \mbox{\epsfxsize=77mm \epsfysize=67mm\epsffile{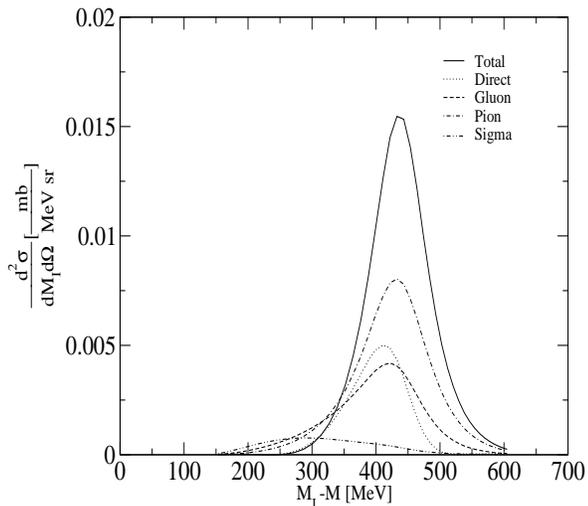}}
\end{center}
\par
\centering
\caption{Detailed contributions to the $p(d,d')$ cross section coming from the
different interactions at the quark level, neglecting the interference
terms. We denote by direct the result obtained neglecting quark-exchange 
diagrams. $M_I$ is the invariant mass of the target system.}
\label{fighire1}
\end{figure}

It is also worth wile to compare our results to the ones obtained from the
baryonic calculation of the RET diagram \cite{17}. To make more clear the
comparison we plot the dependence of these results on the value chosen for
the $\sigma $ mass (within the allowed experimental interval), Fig. 
\ref{hireWID1}, and on the
baryonic cut-off mass needed, Fig. \ref{hireWID2}. 
As can be seen the smaller $m_{\sigma }$ the
bigger the cross section and the smaller the cut-off the smaller the cross
section. The significant dependence on the cut-off mass points out the need
of having a good description of the scalar short-range part of the
interaction. In our quark model framework this scalar piece is not
uncertainly dependent on any free parameter but determined by quark
antisymmetry plus the dynamics, the OPE and OGE giving most of the cross
section, see Fig. \ref{fighire1}. 
It is then clear that the results obtained
with the quark-model derived interactions are
qualitatively quite different to the ones reported using baryonic degrees of
freedom. In fact, the baryonic form-factor could be hiding the effects of the
quark substructure that we find in our quark-model treatment through the
contributions to the scalar channel from every term in the quark-quark
Hamiltonian. 

\section{Summary}

We have carried out a test of a quark-model based
$NN\rightarrow NN^{\ast }(1440)$
potential derived from an universal $qq$ interaction. The consideration of the
long-range tail of the potential as compared to the baryonic parametrization
allows the extraction of the $\pi NN^{\ast }(1440)$ 
and $\sigma NN^{\ast }(1440)$
coupling constants. On the other hand the consideration of the physical
mechanisms involved in the reactions $p(\alpha ,\alpha ^{\prime })$ and 
$p(d,d^{\prime })$, in particular the RET, allows to test the scalar
short-range part of the interaction. The results we get are quite
encouraging in spite of the lack of a full quark model calculation. To
pursue this could open a new way to search for effects of the microscopic
structure in the mentioned processes.

\section{Acknowledgments}

The authors thank Dr. S. Hirenzaki for useful correspondence
concerning the calculation of Ref. \cite{17}.
This work has been partially funded by Ministerio 
de Ciencia y Tecnolog{\'\i}a
under Contract No. BFM2001-3563, by Junta de Castilla y Le\'{o}n under
Contract No. SA-109/01, and by EC-RTN (Network ESOP) under Contracts No.
HPRN-CT-2000-00130 and HPRN-CT-2002-00311.

\end{document}